\documentstyle[12pt,graphics,aasms4]{article}

\newcommand{\PREP}[1]{}

\newcommand{\FF}{\mbox{\bf F}}
\newcommand{\GG}{\mbox{\bf g}}
\newcommand{\II}{\mbox{\bf I}}
\newcommand{\ee}{\hat{{\bf e}}_r}
\newcommand{\BB}{\mbox{\bf B}}
\newcommand{\vv}{\mbox{\bf v}}

\begin{document}
\title{Stellar winds, dead zones, and coronal mass ejections}
\author{R. Keppens and J. P. Goedbloed}
\affil{FOM-Institute for Plasma-Physics Rijnhuizen, 
P.O. Box 1207, 3430 BE Nieuwegein, The Netherlands}
\authoremail{keppens@rijnh.nl}

\begin{abstract}
Axisymmetric stellar wind solutions are presented,
obtained by numerically solving the ideal magnetohydrodynamic (MHD) equations.
Stationary solutions are critically analysed using the knowledge of
the flux functions. These flux functions enter 
in the general variational principle governing
all axisymmetric stationary ideal MHD equilibria.

The magnetized wind solutions for (differentially) rotating stars contain
both a `wind' and a `dead' zone. We illustrate the influence of
the magnetic field topology on the wind acceleration
pattern, by varying the coronal field strength and the
extent of the dead zone. This is evident from the resulting variations 
in the location and appearance of the critical curves where the wind
speed equals the slow, Alfv\'en, and fast speed. Larger dead zones cause
effective, fairly isotropic acceleration to super-Alfv\'enic
velocities as the polar, open field lines are forced to fan out
rapidly with radial distance. A higher field strength moves the
Alfv\'en transition outwards. In the ecliptic, the wind outflow is 
clearly modulated by the extent of the dead zone.

The combined effect of a fast stellar rotation and an equatorial `dead' zone
in a bipolar field configuration can lead to efficient thermo-centrifugal
equatorial winds. Such winds show both a strong
poleward collimation and some equatorward 
streamline bending due
to significant toroidal field pressure at mid-latitudes.
We discuss how coronal mass ejections are
then simulated on top of the transonic outflows.
\end{abstract}


\section{Introduction}\label{s-intro}

The solar wind outflow presents a major challenge to numerical
modeling since it is a fully three-dimensional (3D), time-dependent 
physical environment, where regions of supersonic and subsonic
speeds coexist in a tenuous, magnetized plasma. 
Ulysses observations (McComas et al.~\cite{swoops})
highlighted again
that the solar wind about the ecliptic plane is fundamentally
dynamic in nature, while the fast speed wind across both solar poles
is on the whole stationary and uniform. 
Recent SOHO measurements (Hassler et al.~\cite{hassler})
demonstrated how the fast wind emanating from coronal holes is rooted to the
`honeycomb' structure of the chromospheric magnetic network, making the outflow
truly 3D, while the daily coronal mass ejections are in essence
highly time-varying. Moreover,
one really needs to study these time-dependent, multi-dimensional
aspects in conjunction with the coronal heating puzzle
(Holzer \& Leer~\cite{holzer}). 

Working towards that goal, Wang et al. (\cite{wang}) recently modeled the
solar wind using a two-dimensional, time-dependent,
magnetohydrodynamic (MHD) description with heat and momentum addition as
well as thermal conduction. Their magnetic topology shows
both open (polar) and closed (equatorial) field line regions.
When heating the closed field region, a sharp streamer-like
cusp forms at its tip as the region continuously expands and evaporates.
A quasi-stationary wind model results where the emphasis is on reaching a
qualitative and quantitative agreement with the 
observed latitudinal variation (reproducing, in particular, 
the sharp transition at roughly $\pm 20^\circ$ latitude between fast and
slow solar wind) by tuning the spatial dependence of the
artificial volumetric heating and momentum sources. 

We follow another route towards global solar wind modeling, working our way
up stepwise from stationary 1D to 3D MHD configurations. In a
pure ideal, stationary, axisymmetric MHD approach, numerical simulations
can benefit greatly from analytical theory. This is demonstrated in
Ustyugova et al. (\cite{love}), where stationary magneto-centrifugally driven
winds from rotating accretion disks were calculated numerically and 
critically verified by MHD theory. 

In this paper, we extend the Wang et al. (\cite{wang}) modeling efforts
to 2.5D, by including toroidal vector components while remaining
axisymmetric. This allows us to explore stellar wind regimes where
rotation is also important. The magnetic field still has open and
closed field line regions, but in ideal MHD, the closed field region
is a `dead' zone from where no plasma can escape. The unknown coronal
heating is avoided by assuming a polytropic equation of state and 
dropping the energy equation all together. The stationary, axisymmetric,
polytropic MHD models are analysed as in Ustyugova et al. (\cite{love}).

In particular, we investigate the effects of (i) having both open and
closed field line regions in axisymmetric stellar winds; and of (ii)
time-dependent perturbations within these transonic outflows. 
While we still ignore the basic question of why there should be a hot
corona in the first place, we make significant progress 
towards fully 3D, dynamic models. The advantages of a gradual
approach towards such `final' model were pointed out in 
Keppens \& Goedbloed (\cite{wind98}). 
There, we initiated our effort to numerically 
model stellar outflows by gradually relaxing the assumptions
inherent in the most well-known solar wind model: the isothermal Parker
wind (Parker~\cite{parker}). 
In a sequence of stationary, 1D, 1.5D, and 2.5D, hydrodynamic and
magnetohydrodynamic stellar wind models, all obtained with
the Versatile Advection Code (VAC, see T\'oth~\cite{vacapjl,vac2};
T\'oth \& Keppens~\cite{vac3};
Keppens \& T\'oth~\cite{vac4}, and
{\tt http://www.phys.uu.nl/}\~{\tt toth}),
we demonstrated that
we can now routinely calculate axisymmetric magnetized wind solutions
for (differentially) rotating stars. An important generalization of
previous modeling efforts (Sakurai~\cite{sakuraiAA,sakurai})
is that the field topology
can have both open and closed field line regions, so we 
model `wind' and `dead' zones self-consistently. 
In essence, our work extends the early model efforts by
Pneuman \& Kopp (\cite{pneu}) in (i) going from an isothermal to a
polytropic equation of state; (ii) allowing for stellar rotation;
and (iii) including time-dependent phenomena. While we get qualitatively
similar solutions for solar-like conditions, we differ entirely in
the numerical procedure employed and in the way boundary conditions
are specified. Keppens \& Goedbloed~(\cite{wind98}) contained one such
MHD model for fairly solar-like parameter values.
In this paper, we start with a critical examination of 
this `reference' model. The obtained transonic outflow,
accelerating from subslow to superfast speeds, must obey the conservation
laws predicted by theory, by conserving various physical quantities
along streamlines. This will be checked in Sect.~\ref{s-cons}.
Section~\ref{s-ext} continues with a physical analysis of the
model and investigates the influence of the magnetic field
strength and of the latitudinal extent of the `dead' zone. These
parameters have a clear influence on the global
wind structure, especially evident in the appearance and location of
its critical surfaces where the wind speed equals the slow, Alfv\'en,
and fast magnetosonic speeds. We also present one such wind solution for
a star which rotates twenty times faster than our sun.
Finally, Sect.~\ref{s-cme} relaxes the stationarity of the
wind pattern, by forcing coronal mass ejections on top of
the wind pattern. Conclusions are given in Sect.~\ref{s-conc}.

\section{Reference model and conservation laws}\label{s-cons}

\subsection{Solution procedure}
We recall from Keppens \& Goedbloed~(\cite{wind98}) that we solve
the following conservation laws
for the density $\rho$, the momentum vector $\rho \vv$,
and the magnetic field $\BB$:
\begin{equation}
\frac{\partial \rho}{\partial t}+\nabla \cdot (\rho \vv)=0,
\label{q-mass}
\end{equation}
\begin{equation}
\frac{\partial (\rho \vv)}{\partial t}+ \nabla \cdot [ \rho \vv \vv + p_{tot} \II- \BB \BB]=\rho \GG,
\label{q-mom}
\end{equation}
\begin{equation}
\frac{\partial \BB}{\partial t}+ \nabla \cdot (\vv \BB-\BB \vv)= 0.
\label{q-b}
\end{equation}
Here, $p_{tot}=p + \frac{1}{2}B^2$ is the total pressure,
$\II$ is the identity tensor, $\GG=-(G M_*/r^2)\ee$ is the
external stellar (mass $M_*$) gravitational field with
$r$ indicating radial distance.
We assume $p\sim\rho^{\gamma}$
(dimensionless, we take $p=\rho^\gamma/\gamma$), 
where in this paper we
only construct models for specified polytropic index $\gamma=1.13$.
This compares to the value 1.05 used in recent
work by Wu, Guo, \& Dryer~(\cite{wusp97}) and an empirically determined
value of 1.46 derived from Helios~1 data by Totten, Freeman, \&
Arya~(\cite{totten}).  

The discretized Eqs.~(\ref{q-mass})--(\ref{q-b}) are solved on a
radially stretched polar grid in the poloidal plane using
a Total Variation Diminishing Lax-Friedrich discretization 
(see e.g. T\'oth \& Odstr\v cil~\cite{vac1})
with Woodward limiting (Collela \& Woodward~\cite{wood}). 
Stationary ($\partial/\partial t= 0$) solutions are identified
when the relative change in the conservative variables from
subsequent time steps drops below a chosen tolerance (sometimes down
to $10^{-7}$).
We explained in Keppens \& Goedbloed~(\cite{wind98}) 
how we benefitted greatly from
implicit time integration (see T\'oth, Keppens, \& Botchev~\cite{implvac2};
Keppens et al.~\cite{implvac1}; van der Ploeg, Keppens, T\'oth~\cite{auke}) 
for obtaining axisymmetric 
($\partial/\partial \varphi=0$)
hydrodynamic ($\BB=0$)
stellar outflows characterized by $\rho(R,Z)$ and $\vv(R,Z)$, where $(R,Z)$ are
Cartesian coordinates in the poloidal plane. Denoting
the base radius by $r_*$,
these hydrodynamic models cover $r \in [1,50]r_*$
and have as escape speed $v_{\rm esc}=\sqrt{2GM_*/r_*}=3.3015 c_{s*}$, 
with $c_{s*}$ the base sound speed.
They are also characterized by a rotational parameter
$\zeta=\Omega_* r_*/c_{s*}=0.0156$ (if not specified otherwise), and
impose boundary conditions at the base such that (i) $v_{\varphi}=\zeta R$;
and (ii) the poloidal base speed $\vv_{p}$
is in accordance with
a prescribed radial mass flux $\rho \vv_{p}= f_{\rm mass}\ee/r^2$.
The value of the mass loss rate parameter
$f_{\rm mass}$ is taken from a 1D polytropic,
rotating Parker wind valid for the equatorial regions under identical
parameter values. For $\zeta=0.0156$, we get $f_{\rm mass}=0.01377$.
We clarify below in which way the values for the dimensionless 
quantities $v_{\rm esc}/c_{s*}$, $\zeta$, and
$f_{\rm mass}$ relate to the prevailing solar conditions.

To arrive at a `reference' MHD wind solution, two more parameters
enter the description which quantify the initial field strength and
the desired extent of the `dead' zone. A stationary, axisymmetric,
magnetized stellar wind is the end result of a time stepping process
which has the initial density $\rho(R,Z)$ and toroidal velocity
component $v_{\varphi}(R,Z)$ from the HD solution with identical
$\gamma$, $v_{\rm esc}$, and $\zeta$ parameters. The poloidal velocity
is also copied from the HD solution in a polar `wind' zone where
$\theta < \theta_{\rm wind}$ (upper quadrant with $\theta=0$ at pole),
quantified by its polar angle $\theta_{\rm wind}$. 
The `dead' zone is appropriately initialized by a zero poloidal
velocity. The field is initially set to a monopole field in the
`wind' zone, where 
\begin{equation}
B_R(R,Z;t=0)=B_0 R/r^3, \qquad B_Z(R,Z;t=0)=B_0 Z/r^3,
\end{equation}
where $r^2=R^2+Z^2$, coupled to a dipole field in the `dead' zone with
\begin{equation}
B_R(R,Z;t=0)=3 a_d \frac{Z\,R}{r^5}, \qquad 
B_Z(R,Z;t=0)= a_d \frac{(2 Z^2 -R^2)}{r^5}.
\end{equation}
The strength of the dipole is taken $a_d=B_0/(2 \cos\theta_{\rm wind})$ to
keep the radial field component $B_r$ continuous at $\theta=\theta_{\rm wind}$.
The initial $B_\varphi$ component is zero throughout.
Keppens \& Goedbloed~(\cite{wind98}) took $B_0=3.69$ and
$\theta_{\rm wind}=60^\circ$, so that the corresponding dead zone
covered only a $\pm 30^{\circ}$ latitudinal band about the stellar
equator. For the sun at minimum activity, the extent of the
coronal hole is typically such that $\theta_{\rm wind}=30^{\circ}$,
so it will be useful to vary this parameter in what follows (Sect.~\ref{s-ext}).
We use a resolution of $300 \times 40$ in the full poloidal
halfplane, and impose symmetry conditions at both poles, and
free outflow at the outer radius $50 r_*$ (where all quantities are
extrapolated linearly in ghost cells). At the stellar base,
we similarly extrapolate density and all magnetic field components
from their initial values, but let these quantities adjust in value
while keeping this initial gradient in the ghost cells. 
This implies that the density and the magnetic field at the base 
is determined during the time stepping process to arrive at steady-state.
We enforce the $\nabla \cdot \BB =0$ condition
using a projection scheme (Brackbill \& Barnes~\cite{barnes}),
to end up with a physically realistic magnetic configuration
(despite the `monopolar' field in the wind zone). 
The stellar boundary condition for the momentum equation allows
us to specify a differential rotation rate $\zeta(\theta)$ and latitudinally
varying mass flux through $f_{\rm mass}(\theta)$. We set
\begin{equation}
\rho \vv_{p}= f_{\rm mass}(\theta)\ee/r^2 , \qquad
v_{\varphi}=\zeta(\theta)R+B_{\varphi}v_{p}/B_{p}.
\end{equation} 
The reference
model has a rigid rotation rate according to $\zeta=0.0156$, while
$f_{\rm mass}=0.01377$ in the wind region and zero in the equatorial dead
zone.

As emphasized in Keppens \& Goedbloed~(\cite{wind98}), our choice of
boundary conditions is motivated by the variational
principle governing all axisymmetric, stationary, ideal MHD equilibria
(see Sect.~\ref{ss-stream}). The analytic treatment
shows that the algebraic Bernoulli equation, together with the cross-field
momentum balance, really determine the density profile and the magnetic flux
function concurrently. In keeping with this formalism, we impose a base mass 
flux and a stellar rotation, and let the density and the magnetic field 
configuration adjust freely at the base. 
In prescribing the stellar rotation, we exploit the freedom available in
the variational principle by setting a flux function at the base.
Noteworthy, the Pneuman \& Kopp
(\cite{pneu}) model, as well as many more recent modeling efforts for
stellar MHD winds, fix the base normal component of the magnetic field together
with the density. Below, we demonstrate that our calculated 
meridional density structure compares well with recent observations
by Gallagher et al.~(\cite{gall}).

The values for the dimensionless parameters
$v_{\rm esc}/c_{s*}$, $\zeta$ and $B_0$ (actually the ratio of the coronal
Alfv\'en speed to $c_{s*}$) are solar-like in the following sense.
At a reference radius $r_*=1.25 R_{\odot}$, we
take values for the number density $N_o\simeq 10^8 \,\,\rm{cm}^{-3}$,
temperature $T_o=1.5\times 10^6 \,\,\rm{K}$, coronal field strength
$B_o\simeq 2 \,\,\rm{G}$,
and rotation rate $\Omega_{\odot}=2.998 \times 10^{-6} \,\,\rm{s}^{-1}$.
For $\gamma=1.13$ and assuming a mean molecular weight $\tilde{\mu}=0.5$, the
base sound speed then turns out to be $c_{s*}=167.241 \,\,\rm{km/s}$, with
all dimensionless ratios as used in the reference model.
Further, the value for the mass loss rate parameter 
$f_{\rm mass}=0.01377$ is then in
units of $1.06\times 10^{13} \,\,\rm{g}/\rm{s}$, so that a split-monopole
magnetic configuration leads to a realistic mass loss rate $\dot{M}
\propto 4\pi f_{\rm mass} \simeq 2.9\times
10^{-14}M_{\odot}{\rm{yr}^{-1}}$. Since the reference model has a constant
mass flux in its wind zone, the presence of the dead zone reduces this
value by exactly $(1-\cos\theta_{\rm wind})=1/2$.
Units enter through the reference radius $r_*$, the base sound speed
$c_{s*}$, and the base density $\rho_{*}=N_o m_p \tilde{\mu}$
(with proton mass $m_p$).

\subsection{Streamfunctions}\label{ss-stream}

The final stationary wind pattern is shown below in Fig.~\ref{f-winds} 
(see also Fig.~5 in Keppens \& Goedbloed~\cite{wind98}). 
The physical correctness of this
numerically obtained ideal MHD solution can be checked as follows.
All axisymmetric stationary
ideal MHD equilibria are derivable from a single variational
principle $\delta L=\delta\int{\cal L} dV =0$ with Lagrangian
density (Goedbloed,
Keppens, Lifschitz~\cite{eps98}; Keppens \& Goedbloed~\cite{soho7}):
\begin{equation}
{\cal L}(M^2,\psi,\nabla\psi;R,Z)=
\frac{1}{2R^2}(1-M^2)\mid\nabla\psi\mid^2
-\frac{\Pi_1}{M^2}+\frac{\Pi_2}{\gamma M^{2\gamma}}-\frac{\Pi_3}{1-M^2}.
\end{equation}
To obtain an analytic ideal MHD solution, the
minimizing Euler-Lagrange equations need to be solved simultaneously
for the poloidal 
flux function $\psi(R,Z)$ and the squared poloidal Alfv\'en Mach number
$M^2(R,Z)\equiv \rho v^2_p/B^2_p$. Here,
$\BB_p = (1/R)\hat{{\bf e}}_{\varphi}\times \nabla \psi$.
In contrast with the translationally symmetric case
(Goedbloed \& Lifschitz~\cite{hans1}; Lifschitz \& Goedbloed~\cite{hans2}), 
the governing
variational principle contains factors $R^2$, while the profiles $\Pi_1$
and $\Pi_3$ are no longer flux functions. In particular,
\begin{equation}
\Pi_1 \equiv \chi'^2\left(H + \frac{R^2\Omega^2}{2} + \frac{GM_*}{r}\right) \,,
\end{equation}
\begin{equation}
\Pi_2 \equiv \frac{\gamma}{\gamma-1}\chi'^{2\gamma}S \,,
\end{equation}
\begin{equation}
\Pi_3 \equiv \frac{\chi'^2}{2}\left(R\Omega - \frac{\Lambda}{R}\right)^2 \,.
\end{equation}
where five flux functions $H,\Omega,S,\Lambda,\chi'$ enter. 
These direct integrals of the axisymmetric, stationary ideal MHD equations
are: 
\begin{itemize}
\item the Bernoulli function ($\sim$ energy)
\begin{equation}
H(\psi) \equiv \frac{1}{2} v^2 + \frac{\rho^{\gamma-1}\gamma S}{\gamma-1}-\frac{G M_*}{r}-v_\varphi^2 +v_{\varphi} B_{\varphi} \frac{v_p}{B_p},
\end{equation}
\item the derivative of
the stream function $\chi'\equiv\partial\chi/\partial\psi$.
Indeed, the poloidal stream function $\chi(R,Z)$ necessarily
obeys $\chi(\psi)$, provided that the toroidal component
of the electric field vanishes. These are immediate checks
on the numerical solution, namely $v_R B_Z = v_Z B_R$, or the fact
that streamlines and field lines in the poloidal plane must be parallel
(easily seen in Fig.~\ref{f-winds}). 
\item the entropy $S$, which for our {\it polytropic}
numerical solutions is constant by construction: $S\equiv 1/\gamma$,
\item a quantity related to the angular momentum flux
$\FF_{\rm AM}=\rho \vv_p R v_{\varphi}- \BB_p R B_{\varphi} 
\equiv \rho \vv_p \Lambda$, defined as
\begin{equation}
\Lambda(\psi)\equiv R v_\varphi - R B_\varphi \frac{B_p}{\rho v_p}, 
\end{equation}
\item  and the derivative of the electric field potential
\begin{equation}
\Omega(\psi) \equiv \frac{1}{R} \left(v_{\varphi} - \frac{v_p}{B_p} B_\varphi \right). 
\end{equation} 
\end{itemize}
Various combinations of these flux functions can be made, for instance
Goedbloed \& Lifschitz (\cite{hans1}) 
used the following flux function (instead of $\Lambda$)
\begin{equation}
I(\psi)\equiv R B_\varphi - R v_\varphi \frac{\rho v_p}{B_p} = -\chi' \Lambda .
\end{equation}
Figure~\ref{f-cons} presents gray-scale contour plots of
three streamfunctions
$I$ (actually $\log\mid I\mid$), $\Lambda$, and $H$, calculated from
the reference solution, with its poloidal field lines overlaid. 
Ideally, these contours must match the field line structure exactly: all
deviations are due to numerical errors. Inspection shows that the
agreement is quite satisfactory. A quantitative measure of the
errors is given in the lower two frames. At bottom left, we plotted
the relative deviation of $\Omega$ from the value enforced at the
base $\zeta(\theta)$ (constant to 0.0156 in this model). The solid
line marks the 10\% deviation, actual values range from [0.0092, 0.0187].
It should be noted that this 
is a very stringent test of the solution, since for the chosen
parameters, the wind is purely thermally driven, and the stellar
rotation is dynamically unimportant. 
The largest deviations are apparent at the rotation axis 
(symmetry axis) in $\Omega$, which
is not unexpected due to its $1/R$ dependence. Other inconsistencies
are in the region which has drastically
changed from its initial zero velocity, purely dipole magnetic field structure:
open field lines coming from the polar regions are now
draped around a dipolar `dead' zone of limited radial extent.
This dead zone simply corotates with the base angular velocity, and
has a vanishing poloidal velocity $v_p$ and toroidal field $B_\varphi$.
Around that zone, 
the stellar wind traces the open field lines. 
The final bottom right frame shows $E_\varphi$, virtually vanishing
everywhere, and only at the very base
are values of order ${\cal{O}}(10^{-2})$. For completeness, we show
the plasma beta $\beta=2p/B^2=1$ contour which
exceeds unity in an hourglass pattern 
that stretches out from the dead zone to large radial distances.

Overall, the obtained 
stationary numerical solution passes all criteria for being physically 
acceptable. We expect that most errors disappear when using
a higher resolution. We already exploited
a radial grid accumulated near the stellar surface, necessary for resolving
the near-surface acceleration. However, we could benefit also from a
higher resolution in polar angle, now only 40 points for the full half circle,
by, for instance, using the up-down symmetry.

\section{Extensions of the reference model}\label{s-ext}

With the accuracy of the numerical solutions confirmed by inspection
of the streamfunctions, we can start the discussion of the influence
of the physical parameters $B_0$, $\theta_{\rm wind}$, and $\zeta$ on the
global wind structure. First, we present a more detailed analysis of
the reference solution itself.

Fig.~\ref{f-new} shows the density structure at left, where we plot 
number density as a function of polar angle for three fixed radial
distances, namely at the base $1.27 R_\odot$, at $11.9 R_\odot$ and 
at $12.7 R_\odot$. Keppens \& Goedbloed (\cite{wind98}) already demonstrated
the basic effect visible here: the equatorial density is higher than
the polar density. A recent determination of the $(r,\theta)$ dependence
of the coronal electron densities within $1 R_\odot \leq r \leq 1.2 R_\odot$
by Gallagher et al. (\cite{gall}) concluded that the density falloff is
faster in the equatorial region than at the poles, and that the equatorial
densities within the observed region are a factor of three larger than in the 
polar coronal hole. Their study lists number densities of order 
$10^8 {\rm cm}^{-3}$, as in our reference model. Our base density 
at $1.27 R_\odot$
(dotted in Fig.~\ref{f-new}, left panel) has a distinct latitude
variation reflecting the combined open-closed field line structure.
Interestingly, the observations in Gallagher et al. (\cite{gall})
show a similar structure, with quoted values of 
$8.3 \times 10^7 {\rm cm}^{-3}$ at $1.2 R_\odot$
above the pole, increasing to $1.6 \times 10^8 {\rm cm}^{-3}$ at the same
distance along the equator. In fact, a dip
in the density variation
was present due to an active region situated above the equator. 
Qualitatively, we recover this variation at the boundary of the
dead zone. 
Again, we stress that the base density is calculated self-consistently,
hence not imposed as a base boundary condition.
The other two radial cuts situated beyond the dead zone agree
quite well with the conclusions drawn by the observational study.

The reference wind solution also conforms with some well-known
studies in MHD wind modeling. Suess \& Nerney (\cite{suess})
and Nerney \& Suess (\cite{nern}) pointed out how consistent axisymmetric
stellar wind modeling which include magnetic fields and rotation
automatically lead to a meridional flow away from
the equator. At large radial distances, the flow profile should
be of the form $v_\theta \propto - \sin (2 \theta)$, with a
poleward collimation of the magnetic field.
This variation in polar angle is general and independent of the precise
base field structure. In the middle panel of Fig.~\ref{f-new}, we show
the latitude dependence of $v_\theta$ at $50 R_\odot$ for the reference
model. Note the perfect agreement with the predicted variation.

Since the solar rotation rate, quantified by the parameter $\zeta$, is
low, the calculated wind solution in the poloidal plane should be 
similar to the one presented by Pneuman \& Kopp (\cite{pneu}). They
constructed purely poloidal,
isothermal and axisymmetric models of the solar wind including
a helmet streamer (or `dead' zone). An iterative technique was used
to solve for the steady coronal expansion, while the density and the
radial magnetic field were fixed at the base. They enforced
a dipolar $B_{r}$ with a strength of $1 {\rm G}$ at the poles, half
the value we use at the initialization. Their uniform coronal 
temperature was taken to be $1.56 \times 10^6 {\rm K}$, almost 
identical to our base temperature $T_o$. Their base number density
was imposed to be $1.847 \times 10^8 {\rm cm}^{-3}$, independent
of latitude, and they assumed a slightly higher value for the
mean molecular weigth, namely $\tilde{\mu}=0.608$. This
leads to a base density which is a factor of 2.246 higher than the
one used in our model. With these differences in mind (together with
our polytropic equation of state and the rotational effects), we
show in the right panel of Fig.~\ref{f-new} the magnetic structure
and the location of the sonic (where
$v_{p}=c_s$) and the Alfv\'enic surface 
(where $v_{p}=B_{p}/\sqrt{\rho}$)
in a manner used in the original publication of Pneuman \& Kopp (\cite{pneu}),
their Fig.~4. In the $(1/r, \cos\theta)$ projection, the Alfv\'enic
transition on the equator is at the cusp of the helmet structure
before the sonic point,
while the sonic surface is closer to the solar surface at the poles.
The qualitative agreement is immediately apparent, although our solution
method is completely different, most notably in the prescription of
the boundary conditions. By calculating the base density and magnetic
field configuration self-consistently, we generalize the solution
procedure employed by Pneuman \& Kopp (\cite{pneu}) as we gain control
of the size of the dead zone through our parameter $\theta_{\rm wind}$. 
This allows us to study the influence of the base topology
of the magnetic field on the global wind acceleration pattern in what follows.

Fig.~\ref{f-winds} confronts three steady-state wind solutions
with our reference model, which differ in the latitudinal extent
of the dead zone and/or in the magnetic field strength.
With $\theta_{\rm wind}= 60^\circ$ and $B_0=3.69$ (corresponding
to a $2 {\rm G}$ base coronal field strength) 
for the reference case A, we increased
the latitudinal extent of the dead zone by taking 
$\theta_{\rm wind}=30^\circ$ in
case B, doubled the field strength parameter $B_0$ in model C, 
and took both $B_0=7.4$ and $\theta_{\rm wind}=30^\circ$ to arrive at model D. 
We recall that $B_0$ specifies only the initial field strength used in the
time-stepping process towards a stationary solution. The final base field
strength turns out to be of roughly the same magnitude, 
but differs in its detailed
latitudinal variation. The changes in the gobal wind pattern are 
qualified by the resulting deformations of the critical surfaces 
(hourglass curves) where
the wind speed equals the slow, Alfv\'en, and fast speeds. The plotted
region stretches out to $\approx 18\, r_* \sim 22.5 R_\odot$.

By enlarging the dead zone under otherwise identical conditions
(from A to B), the polar, open field lines are forced to fan out
more rapidly with radial distance. As a result, the acceleration
of the plasma occurs closer to the stellar surface, and the
critical curves become somewhat more isotropic in polar angle. The Alfv\'en
surface moves inwards at the poles, and shifts outwards above the now
larger dead zone at the equator, approaching a circle with
an equatorial imprint of the dead zone. 
If we keep the dead zone small, but double the initial field strength
$B_0$ (from A to C), the opposite behaviour occurs: the critical curves,
hence the entire acceleration behaviour of the wind, become much 
more anisotropic. The most pronounced change is an inward shift
of the polar slow transition, and an outward shift of the Alfv\'en
and fast polar transition. This behaviour is in agreement with
what a Weber-Davis model (Weber \& Davis~\cite{wd})
predicts to happen when the
field strength is increased (note that the Weber-Davis model
only applies to the equatorial region).
When both the field strength and the dead zone are doubled (from
A to D), the resulting Alfv\'en and fast critical curves are rather isotropic
due to the influence of the dead zone. The polar slow transition
is displaced inward while the polar Alfv\'en and fast curve
are shifted outward, as expected for the higher field. 
The detailed equatorial behaviour is clearly modulated by
the existing dead zone.
We note that all wind solutions presented are still thermally
driven, since the solar-like rotation rate is rather low
and the field strengths are very modest. The changes are
entirely due to reasonable variations in magnetic field
topology and only a factor of two in field strength. One could tentatively
argue that such variations occur 
in the solar wind pattern within its 11-year magnetic cycle.
In gray-scale, Fig.~\ref{f-winds} shows the absolute value of
the toroidal field component $\mid{B_\varphi}\mid$ (this field changes sign
across the equator). Note that the stellar rotation has wound up
the field lines in a zone midway between the poles and the
dead zone. For higher rotation rates (see below), the associated 
magnetic pressure build up due to rotation can influence the wind
pattern and cause collimation (Trussoni, Tsinganos, \& Sauty~\cite{trusso}). 
Due to the four-lobe structure, one
can expect parameter regimes which lead to both poleward collimation
(as in the monopole-field models of Sakurai~\cite{sakuraiAA})
and equatorward streamline bending. 

Figure~\ref{f-cuts} compares the radial dependence of the
poloidal velocity for the four models (A, B, C, D) at the pole (left
panel) and the equator (middle panel). 
For comparison, we overplotted the same quantities in each panel for a solution
with a split-monopole base field at the same parameter values.
This monopolar field solution is identical in nature to the 
Sakurai~(\cite{sakuraiAA,sakurai}) models and was shown in Keppens \& Goedbloed
(\cite{wind98}, Fig.~4). 
At the pole, a faster acceleration
to higher speeds as compared to the reference model results from either
increasing the field strength or enlarging the dead zone. Moreover,
all four
models show a faster initial acceleration than the corresponding monopolar
field model. 
Radio-scattering measurements of the polar solar wind speed 
(Grall et al.~\cite{grall}) indicated that
the polar wind acceleration is almost complete by $10\, R_{\odot}$, much
closer than expected. Our model calculations show that a fast acceleration
can result from modest increases in the coronal field strength and dead
zone extent (model D has a solar-like dead zone of $\pm 60^{\circ}$).

The middle panel of Fig.~\ref{f-cuts}
shows the distinct decrease in equatorial wind speed due to the
dead zone, when compared to a split-monopole solution.
The equatorial velocities are reduced by $10$ to $40 \,{\rm km/s}$, depending
on the size of the dead zone and the base field strength.
Enlarging the dead zone reduces the wind speed significantly
(compare A to B, and C to D).
To a lesser degree, the same effect is true for an increase in
coronal field strength (compare A to C, and B to D). 
Keppens \& Goedbloed~(\cite{wind98}, Figure~6) 
contained a polar plot of the velocity and the density at fixed
radial distance for
the reference model, where at least qualitatively, a transition from
high density, low speed equatorial wind to lower density, high speed polar
wind is noticable. As evidenced by Fig.~\ref{f-cuts}, this difference
in equatorial and polar wind is even more pronounced for larger dead zones.
The velocities reached are too low for explaining the solar wind
speeds -- the Weber-Davis wind solution of identical parameters reaches 
$263\, {\rm km/s}$ at 1 AU.
However, this is a well-known shortcoming of a polytropic MHD description
for modeling the solar wind. Wu et al. (\cite{wujgr99}) therefore resort
to an {\it ad hoc} procedure where the polytropic index is an increasing
function of radial distance, $\gamma(r)$, to attain a more
realistic $420\, {\rm km/s}$ wind speed at 1 AU, corresponding to
the `slow' solar wind. A more quantitative
agreement with the observations at these distances must await models
where we take the energy equation into account and/or model extra
momentum addition as in Wang et al.~(\cite{wang}).
The equatorial toroidal velocity profile is shown at right in 
Fig.~\ref{f-cuts}. Note that
increasing $B_0$ or enlarging the dead zone both negatively affect the
degree to which the corona corotates with the star. 

Keppens \& Goedbloed~(\cite{wind98})
also contained a hydrodynamic solution for a much faster
rotation rate quantified by $\zeta=0.3$, or twenty times the
solar rotation rate. 
The additional centrifugal acceleration moves the
sonic transition closer to the star along the equator, and induces
an equatorward streamline bending at the base at higher latitudes
(see also Tsinganos \& Sauty~\cite{tsinhd}).
One could meaningfully ask what remains of this effect when a
two-component field structure is present as well. Therefore, we
calculate an MHD wind for this rotation rate, with
$B_0=3.69$ and $\theta_{\rm wind}=60^\circ$ as in the reference case.
The corresponding mass loss rate parameter (only used in the wind
zone) is $f_{\rm mass}=0.01553$. 
Fig.~\ref{f-rot} displays the
wind structure as in Fig.~\ref{f-winds}, with the gray-scale
indicating the logarithm of the density pattern.
For this solution, we used the up-down symmetry to double
the resolution in polar angle at the same computational cost.
The shape of the critical curves has changed dramatically,
with a significant outward poleward shift of the Alfv\'en curve,
and a clear separation between Alfv\'en and fast critical
curves -- in agreement with a 1.5D Weber-Davis prediction.
The actual position of the critical surfaces
may be influenced by an interaction with
the outer boundary at $50\, r_*$ in the time-stepping towards a 
stationary solution. In fact, the combined Alfv\'en-fast polar transition
has shifted outside the computational domain, and 
the residual could not be decreased to
arbitrary small values but stagnated at ${\cal{O}}(10^{-7})$ following
this interaction. Within the
plotted region of $\sim 37 R_\odot$, the solution is acceptable
as explained in Sect.~\ref{s-cons}. Note how the density structure shows
an increase towards the equator, causing a very effective thermo-centrifugal
acceleration of the equatorial wind above the dead zone.
The equatorward streamline bending occuring in the purely hydrodynamic
wind is still important, but now clearly affected by
the presence of the dead zone. The toroidal magnetic pressure built up
by the stellar rotation along
the mid-latitude open field lines is shaping the wind structure as a whole.
In those regions, we have $\rho v_{p}^2 / B^2_{\varphi} < 1$ together
with $2 p/B^2_{\varphi} < 1$. 
Thereby, it also leads to streamline bending, both poleward as clearly
seen in the high latitude field lines, and equatorwards in the vicinity
of the stellar surface. 
In this way, the magnetic topology consisting
of a dead and a wind zone, combined with fast rotation, leads to
magnetically dominated collimation along the stellar poles, together
with magneto-rotational deflections along the equator. The latter leads
to enhanced densities in the equatorial plane.

Figure~\ref{f-rotcuts} shows the radial dependence at a
polar angle $\theta=41.6^\circ$ of the radial velocity
$v_r$, sound speed $c_s$, radial Alfv\'en speed $A_r=B_r/\sqrt{\rho}$,
and azimuthal speed $v_\varphi$. Note that the radial velocity
reaches up to $400 \, {\rm km/s}$ (compare with the $\sim 200 \, {\rm km/s}$
velocities reached under solar conditions as shown in Fig.~\ref{f-cuts}),
as a result of the additional centrifugal acceleration.
The rotation plays a significant dynamical role here, in contrast to the
`solar-like' models discussed earlier and displayed in Fig.~\ref{f-winds}. 
The toroidal speed $v_\varphi$ reaches above $100 \,{\rm km/s}$, a factor
of $10$ higher than along the ecliptic as shown in Fig.~\ref{f-cuts}
(left panel). The corotation obtained by the numerical procedure
to find the stationary state can again be quantified by the relative error
$\mid \Omega - \zeta \mid/\zeta$: the bottom panel of
Fig.~\ref{f-rotcuts} proves that it is less than 3 \% along that
same radial cut.

\section{Triggering coronal mass ejections}\label{s-cme}

In the process of generating a stationary wind solution, various
dynamic phenomena take place which may have physical relevance.
For instance, the equatorial conic section delineated by $\theta \in 
[\theta_{\rm wind}, \pi-\theta_{\rm wind}]$
which was initially static ($v_p=0$) and dipolar throughout, first gets
`invaded' by plasma emanating from the wind zone. The open field structure
is dragged in towards the equator, and most of the dipolar field
is moved out of the domain, except for the remaining `dead' zone. One could
qualitatively relate some of these changes in the global magnetic topology
with observed coronal phenomena. 

In reality however, coronal mass ejections represent
major disturbances which happen on top of the stationary transonic solar wind.
They are associated with sudden, significant mass loss and cause violent
disruptions of the global field pattern. Most notably,
one frequently observes the global coronal wind structure to return
to its previous stationary state, after the passage of the CME. Within
the realms of our stellar wind models, we can trigger CMEs on top of
the outflow pattern, study their motion, and at the same time demonstrate
that the numerical solutions indeed are stable to such violent perturbations
by returning to a largely unchanged stationary state.
We still restrict ourselves to axisymmetric calculations, so the geometry
of our `CME' events is rather artificial. In future work, we intend
to model these CMEs in their true 3D setting. 

As background stellar wind, we use a slightly modified
model B from the previous section. Model B had a large dead zone,
$\theta_{\rm wind}=30^{\circ}$, $B_0=3.69$, and a rigid rotation
with $\zeta=0.0156$ corresponding to 
$\Omega_{\odot} \sim 3 \times 10^{-6} \,{\rm s}^{-1}$. We changed the boundary 
condition on $v_\varphi$
to mimick a `solar-like' differential rotation, by taking
\begin{equation}
\zeta(\theta)=\zeta_0 + \zeta_2 \cos^2\theta + \zeta_4 \cos^4\theta,
\end{equation}
with $\zeta_0=0.0156$, $\zeta_2=-0.00197$, and $\zeta_4=-0.00248$.
This enforces the equator to rotate faster than the poles 
in accord with the observations. As expected
for this low rotation rate, this has no significant influence on the
wind acceleration pattern. The coronal mass ejection is an equally
straigthforward modification of the boundary condition imposed
on the poloidal momentum equation, namely
$\rho \vv_p = f_{\rm mass}(\theta,t)\ee/r^2$ with
\begin{equation}
f_{\rm mass}(\theta,t)=f_{\rm wind}(\theta)+ g_{\rm CME} \sin\left(\frac{\pi t}{\tau_{\rm CME}}\right) \cos^2\left(\frac{\pi}{2}\frac{\theta-\theta_{\rm CME}}{a_{\rm CME}}\right),
\end{equation}
for $0\leq t\leq\tau_{\rm CME}$ and $\theta_{\rm CME}-a_{\rm CME} \leq \theta \leq
\theta_{\rm CME}+a_{\rm CME}$, and otherwise 
\begin{equation}
f_{\rm mass}(\theta,t)=f_{\rm wind}(\theta).
\end{equation}
The wind related mass loss rate $f_{\rm wind}(\theta)$ contains the polar angle
dependence due to the dead zone, as before. The extra four
parameters control the magnitude of the CME mass loss rate
$g_{\rm CME}$, the duration $\tau_{\rm CME}$,
and the location $\theta_{\rm CME}$ and extent $0\leq a_{\rm CME}\leq \pi/2$ 
in polar angle for
the mass ejection. We only present one CME scenario for parameter values
$g_{\rm CME}=2$, $\tau_{\rm CME}=0.5$, $\theta_{\rm CME}=60^{\circ}$, and 
$a_{\rm CME}=30^{\circ}$. 
Note that the up-down symmetry is hereby deliberately broken.
This scenario mimicks a mass ejection which detaches from
the coronal base within 45 minutes and which has an associated mass flux of 
about $2\times 10^{13} \,\,\rm{g}/\rm{s}$. 
In fact, the total amount of mass lost due to the CME can be evaluated
from
\begin{equation}
M_{\rm lost}^{\rm CME}=2 g_{\rm CME} \tau_{\rm CME} 
 \frac{\pi^2}{\pi^2 - a_{\rm CME}^2}\left\{\cos(\theta_{\rm CME} - a_{\rm CME})
            -\cos(\theta_{\rm CME} + a_{\rm CME})\right\}.
\end{equation}
For the chosen parameter values, this works out to be 
$M_{\rm lost}^{\rm CME}=\frac{36}{35}\sqrt{3}$, corresponding to 
$0.98 \times 10^{17} {\rm g}$, a typical value for a violent event.

Figure~\ref{f-cme1} shows the density difference between the evolving mass 
ejection and the background stellar wind at left,
the magnetic field structure (middle), and the toroidal velocity component
$v_{\varphi}$ (right) at times $t=1$ (1hr 27' after onset)
and $t=3$ (4hr 20' after onset). The region is plotted up to $15\, r_*\simeq
18.75 R_\odot$. Although the event is triggered
in the upper quadrant dead zone only, its violent character also 
disturbs the overlying open field (or wind) zone. The added plasma, trapped
in the dead zone, even perturbs the lower quadrant wind zone at later
times. Note that the CME induces global, abrupt changes in the
toroidal velocity component.
The outermost closed field lines get stretched out radially, pulling
the dead zone along (see Fig.~\ref{f-cme1}-\ref{f-cme2}). 
In ideal MHD calculations, they can never detach through
reconnection, although numerical diffusion can cause it to happen. We observed
the outermost field lines of the dead zone to travel outwards without
noticable reconnection. The overall wind pattern in the first 15
base radii thereby approximately
returns to its original stationary state, as shown
in Figure~\ref{f-cme2} which gives the solution at $t=5$ (7hr 13' after onset)
and $t=30$ (more than 43 hrs after onset). 
Figure~\ref{f-cmeis} shows how a hypothetical spacecraft at $21r_*\sim
26.23 R_\odot$,
close to the ecliptic, would record the CME-passage as a sudden
increase in density and poloidal velocity which eventually relax to 
their pre-event levels. The event is followed by
an increased azimuthal flow regime and shows large amplitude
oscillations in magnetic field strength and orientation.

As we assumed axisymmetry, this simulation serves
as a crude model for CME-type phenomena. Interestingly, axisymmetric
numerical simulations 
of toroidal flux `belts' launched from within the dead zone of
a purely meridional, polytropic MHD wind can relate favourably
to satellite magnetic cloud measurements at 1AU
(Wu et al.~\cite{wujgr99}). Note that we triggered
a `CME' by prescribing a time and space dependent mass flux at the
stellar base, where the density and the magnetic field components
could adjust freely. Alternatively, as used in studies by Miki\'{c} \&
Linker~(\cite{mikic}), global coronal restructuring can be triggered
by shearing a coronal arcade.
Parameter studies of axisymmetric,
but ultimately 3D solutions, could investigate the
formation and appearance of various MHD shock fronts depending on
plasma beta, Mach numbers, etc.

\section{Conclusions and outlook}\label{s-conc}

Continuing our gradual approach towards dynamic stellar wind
simulations in three dimensions, we studied the influence of
the magnetic field strength and topology (allowing for 
wind and dead zones), of the stellar rotation, and of sudden
mass ejecta on axisymmetric MHD winds. 

We demonstrated how reasonable changes in the coronal magnetic field
(factor of two in field strength and in dead zone extent) influence the
detailed acceleration behaviour of the wind. Larger dead zones cause
effective, fairly isotropic acceleration to super-Alfv\'enic  
velocities since the polar, open field lines are forced to fan out
rapidly with radial distance. The Alfv\'en transition moves
outwards when the coronal field strength increases.
The equatorial wind outflow is in these
models sensitive to the presence and extent of the dead zone, but
has, by construction, a vanishing $B_{\varphi}$ and a
$\beta > 1$ zone from the tip of the dead zone to large radial distances. 
The parameter values for these models are solar-like, hence the winds
are mostly thermally driven and, in particular, emanated from slowly
rotating stars. 

For a twenty times faster than solar rotation rate, the wind structure changes
dramatically, with a clear separation of the Alfv\'en and fast
magnetosonic critical curves. At these rotation rates, a pure
hydrodynamic model predicts equatorward streamline bending from
higher latitudes. Our MHD models show how this is now mediated by
the dipolar dead zone. An equatorial belt of
enhanced density stretches from above the dead zone outwards, where
effective thermo-centrifugally driven outflow occurs. The magnetic field
structure shows signs of a strong poleward collimation, due to
the significant toroidal field pressure build up at these spin rates.
As pointed out by Tsinganos \& Bogovalov (\cite{tsing99}), this situation
could apply to our own sun in an earlier evolutionary phase. 

It could be of interest to make more quantitative parameter
studies of the interplay between field topology, rotation rate, etc. in
order to apprehend transonic stellar outflows driven by combinations
of thermal, magnetic, and centrifugal forces.
A systematic study of the angular momentum loss rates as a function
of dead zone extent, magnetic field strength, and rotation rate
can aid in stellar rotational evolution modeling 
(Keppens, Charbonneau, \& McGregor~\cite{rotpaper}; 
Keppens~\cite{binary}). Specifically, Li (\cite{jianke}) pointed out 
that the present solar magnetic breaking rate is consistent with
either one of two magnetic topologies: (i) one with the standard coronal
field strength of $\sim$ 1 $G$ and a small $< 2 \, R_{\odot}$ dead zone;
or (ii) one with a larger $\sim$ 5 $G$ dipole strength and a sizeable dead zone.
When we calculate the torque exerted on the star by the magnetized winds
A, B, C and D shown in Fig.~\ref{f-winds} as
\begin{equation}
\tau_{\rm wind}=4\pi \int_{0}^{\frac{\pi}{2}}d\theta \, \Lambda \rho r^2 v_R,
\label{q-tor}
\end{equation}
(noting the axi- and up-down symmetry), we find 
$\tau_{\rm wind}^{A}\simeq 0.139 \times 10^{31} {\rm dyne\, cm}$, 
$\tau_{\rm wind}^{B}\simeq 0.062 \times 10^{31} {\rm dyne\, cm}$, 
$\tau_{\rm wind}^{C}\simeq 0.246 \times 10^{31} {\rm dyne\, cm}$, 
and $\tau_{\rm wind}^{D}\simeq 0.123 \times 10^{31} {\rm dyne\, cm}$.
This confirms Li's result, since a simultaneous doubling of
the coronal field strength and the dead zone extent (from model A to D)
hardly changes the torque magnitude. As could be expected,
only enlarging the dead zone lowers the breaking efficiency (as pointed
out in Solanki, Motamen, \& Keppens~\cite{samipap}), while
only raising the field strength leads to faster spin-down. Interestingly,
a Weber-Davis prediction with governing parameters identical to
the reference model A (as presented in Keppens \& Goedbloed~\cite{wind98})
gives a value $\tau_{\rm wind}=
4\pi\rho_A r_A^2 v_{rA} \frac{2}{3}\Omega_* r_A^2 \simeq
2.387 \times 10^{31} {\rm dyne\, cm}$ 
(all quantities evaluated at the Alfv\'en radius $r_A$), 
one order of magnitude larger! The same conclusion was reached
by Priest \& Pneuman (\cite{priest74}) by estimating the angular momentum
loss rate from the purely meridional Pneuman \& Kopp (\cite{pneu})
model. Although the latter model does not include rotation (so that $\Lambda$
in Eq.~(\ref{q-tor}) is strictly zero for this model), 
Priest \& Pneuman (\cite{priest74})
could estimate the torque for a solar rotation rate 
from the obtained variation of the poloidal
Alfv\'en radius as a function of latitude (our Fig.~\ref{f-new}, right panel).
The resulting estimate was only 15~\% of that for a monopole base field.
Our exactly evaluated spin-down rates are 2.6~\% to 10.3~\% of a split-monopole
case. The large difference arises due to the
presence of the dead zone and the fact that $B_{\varphi}$ vanishes across the
equator for the wind solutions from Fig.~\ref{f-winds}. Indeed, evaluating
the torque from Eq.~\ref{q-tor} for the monopolar wind solution from Keppens 
\& Goedbloed~(\cite{wind98}, Figure~4) gives $\tau_{\rm wind}\simeq 2.326
\times 10^{31} {\rm dyne\, cm}$, in
agreement with the Weber-Davis estimate.
Hence, it should be clear that full MHD modeling is a useful tool to 
further evaluate and constrain
different magnetic braking mechanisms.

We showed how CME events can be simulated on top of these
transonic outflows. The detailed wind structure is stable to violent
mass dumps, even when ejected in the dead zone. 
Note that we restricted ourselves to axisymmetric perturbations, and
it will be of interest to show whether the axisymmetric solutions
are similarly stable to non-axisymmetric perturbations
(as recently investigated for shocked accretion flows on compact objects
in Molteni, T\'oth, \& Kuznetsov~\cite{molt}).
One could then focus on truly 3D mass ejecta and their parametric dependence
(possibly allow for direct comparison with
LASCO observations of coronal mass ejections), 
or even experiment with unaligned rotation and
magnetic axes. A 3D time-dependent analytic model 
by Gibson \& Low (\cite{gibson}) can
be used as a further check on the numerics. Alternatively, we may decide to
zoom in on (3D) details of the wind structure
at the boundaries of open and closed field line regions or about
the ecliptic plane, to see whether shear flow driven Kelvin-Helmholtz
instabilities (Keppens et al.~\cite{kh2d}; Keppens \& T\'oth~\cite{kh3d}) 
develop in these regions.

\acknowledgments
     The Versatile Advection Code was developed as part of the project on
     `Parallel Computational Magneto-Fluid Dynamics', funded by the
     Dutch Science Foundation (NWO)
     Priority Program on Massively Parallel Computing, and
     coordinated by JPG. Computer time on the Cray C90
     was sponsored by the
     Dutch `Stichting Nationale Computerfaciliteiten' (NCF).
     We thank Keith MacGregor for suggesting to compare
     torque magnitudes for different models and an anonymous referee for
     making several useful comments.

\newpage
\setlength{\topmargin}{-2cm}
\begin{figure}
\begin{center}
\end{center}
\caption{The numerical, ideal MHD stellar wind can be checked
to conserve various quantities along poloidal streamlines and field lines.
For the reference solution, the poloidal field lines (solid) must be isolevels
in the contourplots of (A) the flux function $I$ (plotted is
$\log \mid I \mid$); (B) the flux function $\Lambda$;
and (C) the Bernoulli function $H$. A quantitative error estimate is
shown in panel (D) where we plot $\mid \Omega - \zeta \mid / \zeta$, where the
solid lines delineate dark-shaded regions with deviations $> 10$ \%. 
Panel (E) shows in a contour plot of $E_\varphi$, that this
toroidal electric field component vanishes nearly everywhere. The solid line
in (E) indicates the $\beta =1$ isolevel.
\label{f-cons}}
\end{figure}
 
\begin{figure}
\begin{center}
\end{center}
\caption{Analysis of the reference solar wind solution. Left panel:
Number density from pole to pole at three fixed radial distances: 
$11.9 R_\odot$ (solid), $12.7 R_\odot$ (dashed), and at the base $1.27 R_\odot$
(dotted and scaled to fit on the figure). Middle panel: the meridional velocity
component $v_{\theta}$ as function of polar angle at a fixed $50 R_{\odot}$.
Right panel: the magnetic field configuration and the poloidal sonic (dashed)
and poloidal Alfv\'en (solid) surface as in Pneuman \& Kopp (1971).
\label{f-new}}
\end{figure}

\begin{figure}


\caption{The variation in the detailed wind acceleration pattern due to changes
in the stellar magnetic field. Poloidal cuts with the star 
at the centre contain magnetic field lines (solid) and poloidal flow vectors,
necessarily parallel to the field lines. The hourglass curves indicate
the critical curves for slow (dotted), Alfv\'en (solid), and
fast (dashed) speeds in the wind acceleration. The gray-scale 
contours indicate the absolute
value of the toroidal field component $\mid B_\varphi \mid$. Starting from
the reference solution in (A), we double the extent of the dead zone in (B),
raise the field strength to twice its value in (C), and double both
the dead zone extent and the field strength in (D).
\label{f-winds}}
\end{figure}

\begin{figure}
\caption{For the four solutions shown in Fig.~\ref{f-winds}, we compare
at left: the poloidal velocity at the pole as a function of radius; middle:
the poloidal velocity along the equator; at right: the toroidal velocity
along the equator. In all three panels, the crosses indicate the same
quantity for a solution with a split-monopole base field.
\label{f-cuts}}
\end{figure}

\begin{figure}
\caption{A magnetized wind solution for a star rotating 20 times faster
than our reference `solar' solution. With the star at centre, field lines
and poloidal flow vectors are shown at right, density contours are
given at left, and the critical slow (dotted), Alfv\'en (solid), and
fast (dashed) curves are shown throughout the poloidal cut, stretching
out to $37 R_\odot$.
\label{f-rot}}
\end{figure}

\begin{figure}
\caption{A purely radial cut through the wind solution from Fig.~\ref{f-rot}
at a polar angle of $\theta=41.6^\circ$. Top panel shows the radial
Alfv\'en speed $A_r$, sound speed $c_s$, radial velocity $v_r$ and
azimuthal velocity $v_\varphi$. The bottom panel confirms the
strict corotation achieved: the (flux) function $\Omega$ deviates
less than 3 \% from its fixed base value $\zeta$.
\label{f-rotcuts}}
\end{figure}

\begin{figure}


\caption{A `coronal mass ejection' simulated on top of an axisymmetric
transonic wind. At the times indicated
(1hr 27' and 4hr 21' after the onset), we show poloidal cuts of at left: the
difference in the density pattern between the evolving CME and the original
stationary wind solution; middle: the poloidal field structure; at right:
the toroidal velocity component.
\label{f-cme1}}
\end{figure}

\begin{figure}


\caption{As Fig.~\ref{f-cme1}, times corresponding to 7hr 13' and 43hr
past the onset. Note how the solution 
returns to a state almost identical to the original stationary wind solution.
\label{f-cme2}}
\end{figure}

\begin{figure}
\caption{In situ measurement of a CME passage: 
number density, poloidal velocity, toroidal velocity, poloidal field strength,
and toroidal field as a function of time at a fixed position of $26.23 R_\odot$ 
and an angle of $2.25^\circ$ above the ecliptic.
\label{f-cmeis}}
\end{figure}

\end{document}